\documentclass[onecollarge,natbib]{svjour2}
\bibpunct{[}{]}{,}{n}{}{,} 
\smartqed  
\usepackage{graphicx}
\journalname{Few-Body Systems (APFB2011)}
\begin{document}

\title{\boldmath
Two-Boson Exchange Physics: A Brief Review
} 


\author{Shin Nan Yang
}


\institute{Shin Nan Yang \at
              Department of Physics, National Taiwan University, Taipei 10617, Taiwan \\
              \email{snyang@phys.ntu.edu.tw}           %
}

\date{Received: date / Accepted: date}

\maketitle

\begin{abstract}
Current status of the two-boson exchange contributions to elastic
electron-proton  scattering, both for parity conserving and
parity-violating, is briefly reviewed. How the discrepancy in the
extraction of elastic nucleon form factors between unpolarized
Rosenbluth and polarization transfer experiments can be understood,
in large part, by the two-photon exchange corrections is discussed.
We also illustrate how the measurement of the ratio between
positron-proton and electron-proton  scattering can be used to
differentiate different models of two-photon exchange. For the
parity-violating electron-proton scattering, the interest is on how
the two-boson exchange (TBE), $\gamma Z$-exchange in particular,
could affect the extraction of the long-sought strangeness form
factors. Various calculations all indicate that the magnitudes of
effect of TBE on the extraction of strangeness form factors is
small, though can be large percentage-wise in certain kinematics.

\end{abstract}

\section{Introduction}
\label{intro}

 As the only stable hadron, the study of the structure of the proton provides a
 unique and excellent testing ground of QCD. In the one-photon exchange (OPE) approximation,
 it is well-known that the proton's electric $(G_E)$ and magnetic
 $(G_M)$ form factors (FFs) can be extracted from the reduced differential cross
 section of the electron-proton $\it (ep)$ elastic scattering as one has
 \begin{eqnarray}
 \sigma_R(Q^2,\epsilon)\equiv\frac {d\sigma}{d\Omega_{lab}}\frac{\epsilon(1+\tau)}{\tau\sigma_{Mott}} = G_M^2 +
 \frac{\epsilon}{\tau}G_E^2, \label{diffCr}
 \end{eqnarray}
 where $\tau=Q^2/4M^2,\,\,
 \epsilon^{-1}=1+2(1+\tau)tan^2\theta/2,\,\,
 Q^2=-q^2$ is the momentum transfer squared, $\theta$   the
 laboratory scattering angle, $0\le\epsilon\le 1$, and $\sigma_{Mott}$ is the Mott cross section.
For fixed $Q^2$, varying angle $\theta$, i.e. $\epsilon$, and
adjusting incoming electron energy as needed to plot $\sigma_R$
versus $\epsilon$ will give the FFs, a method often called the
Rosenbluth, or longitudinal-transverse (LT), separation technique.
Experiments employing such a technique over the last half century
all indicate that $R=\mu_pG_E/G_M \sim 1$ for $Q^2<6 $ GeV$^2$ as
often quoted in textbooks.

On the other hand, it has been shown \cite{Akhiezer58} that, again
in the OPE approximation, the ratio $R$ can be accessed in $ep$
scattering with longitudinally polarized electron  by measuring the
polarizations of the recoiled proton parallel $P_l$ and
perpendicular $P_t$ to the proton momentum in the scattering plane,
\begin{eqnarray}
\frac{G_E}{G_M}=-\sqrt{\frac{\tau(1+\epsilon)}{2\epsilon}}\frac{P_t}{P_l}.\label{polR}
\end{eqnarray}
Polarization transfer (PT) experiment of this kind is only possible
recently at JLab. It came as a big surprise that these PT
experiments yield values of $R$ markedly different from 1 in the
same range of $Q^2$ \cite{Jones00}. It prompts  intensive efforts,
both experimentally and theoretically, to understand this apparent
discrepancy as the form factors, which describe the spatial
distribution of the charge and magnetization  inside the proton, are
among the most fundamental observables in hadron physics.

On the experimental side, the first step is to verify this
discrepancy. A new global analysis of the world's cross section data
was carried out \cite{Arrington03}. It found that the great majority
cross sections were consistent and the disagreement with
polarization transfer measurements remains. A set of extremely high
precision measurements of $R$ was soon performed using a modified
Rosenbluth technique \cite{Qattan05}, with the detection of recoil
proton to minimize the systematic uncertainties, and confirmed the
discrepancy.

On the theoretical side, the immediate step taken was to carefully
reexamine the radiative corrections which were known to be as large
as $30\%$ of the uncorrected cross section in certain kinematics. Of
various radiative corrections, only proton-vertex and two-photon
exchange (TPE) corrections contained $\epsilon$ dependence. The
proton-vertex corrections had been investigated thoroughly by
Maximon and Tjon and \cite{MT00} found to be negligible. Realistic
evaluations of the TPE corrections are hence called for to see
whether they are able to resolve this discrepancy. Both hadronic and
partonic model calculations have found that TPE effects can account
for more than half of the discrepancy which we will briefly review
in Sec. 2.

Because TPE effects were found to play an important role in the
parity-conserving electron-proton scattering, it was natural to
speculate that $\gamma Z$-exchange could also play a non-negligible
role in the parity-violating ${\it ep}$ elastic scattering which has
been intensively studied in order to extract the proton strangeness
form factors. Theoretical efforts in this regard will be briefly
reviewed in Sec. 3. Finally, Sec. 4 will be devoted to summary and
discussions on the outlook of the future research on TBE effects, in
particular the TPE effects in the parity-conserving electron-proton
scattering.

\section{Two-photon exchange in elastic electron-proton scattering}
\label{sec:1} In this section, we   discuss (1) the attempts to
evaluate the TPE contributions to ${\it ep}$ scattering, focusing
mostly on two types of model calculations, one hadronic and the
other partonic, (2) possible phenomenological  parametrization of
the TPE effect, and (3) use of $e^+p/e^-p$ ratio and normal
asymmetries as constraints on models.

\subsection{Model calculations of TPE}
\label{sec:2}

\begin{figure}[htbp]
\begin{minipage}[h]{90mm}
\vspace{0pt} \normalsize{In the well-known  work of Mo and Tsai
\cite{MoTsai69}, the radiative corrections were treated in the
soft-photon approximation. In the case of the box and cross-box
diagrams depicted in Fig. \ref{fig:1}, it amounts to evaluating the
integrand within the integration over the four-momentum of the
internal by taking $k=0$ and $k=q$. After the discrepancy between LT
and PT experiments in extracting $R$ is confirmed, Tjon and his
collaborator \cite{Blunden03} immediately set out to treat the TPE
effects rigorously, in a hadronic model including finite size of the
proton but only the elastic nucleon intermediate states. They made
use  of the package FeynCalc \cite{Feyncalc} and LoopTools
\cite{Looptools} which enable them to carry out integrals over
four-momentum $k$ with numerator in the integrand containing power
of $k$ up to 4. Their results for $R$, corrected with}

\end{minipage}
\hfill
\begin{minipage}[h]{60mm}
\vspace{0pt} \raggedright
\begin{center}

\includegraphics[width=0.85\textwidth]{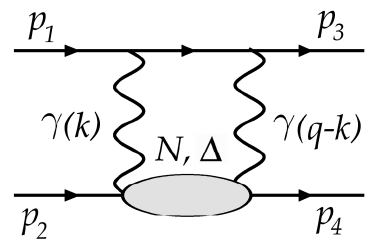}
\end{center}
\caption{Box diagram of the two-photon exchange in electron-proton
scattering with nucleon and $\Delta$ in the intermediate states. The
 cross-box diagram is implied. }
\label{fig:1}       
\end{minipage}
\end{figure}

\begin{figure}[htbp]
\begin{center}
\includegraphics[width=0.45\textwidth]{fig22.eps} \hspace{5mm} \includegraphics[width=0.45\textwidth]{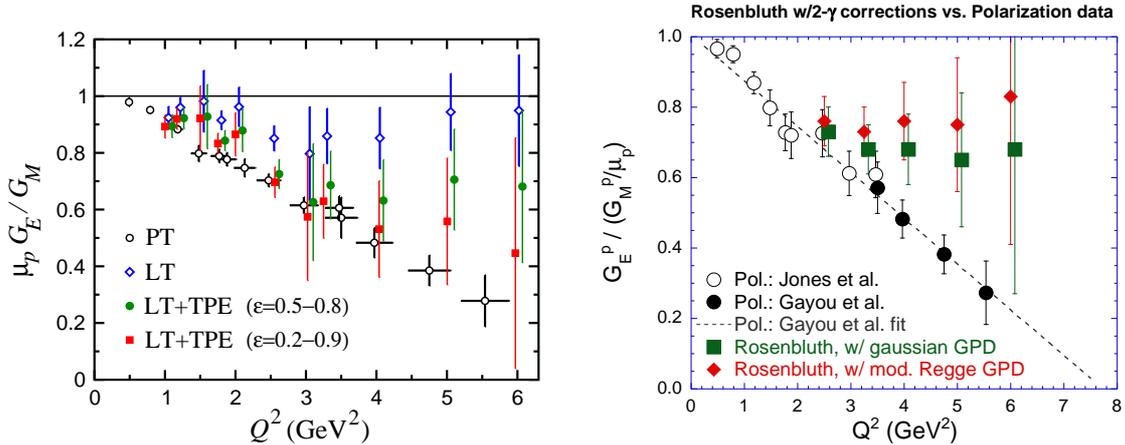}
\end{center}
\caption{TPE corrections to $\mu_p G_E/G_M$. The PT data are from
\cite{Jones00} while the data obtained using LT separation method
are from \cite{Arrington03,Andivahis94}. Left panel: results of the
hadronic model calculation of \cite{Blunden03}, where the figure is
taken. Right panel: results of the partonic  calculation of
\cite{Chen04}, where the figure is taken.
 } \label{fig:2}
\end{figure}

\noindent TPE effects, are shown in the left panel of Fig.
\ref{fig:2} and indicates that TPE coorections can account for more
than half of the discrepancy. Later efforts to include higher
resonances like $\Delta(1232)$ etc. improve the agreement with the
data quantitatively, but not qualitatively.  We refer the readers to
a recent review \cite{Arrington11} for details on these
developments.

The partonic calculation of the TPE corrections was reported in
\cite{Chen04}. The generalized parton distributions which appear in
hard exclusive processes were used to evaluate the handbag diagram
where the incident electron scatters  from quarks in the proton. The
results of this partonic calculation are shown on the right panel of
Fig. \ref{fig:2} which are qualitatively similar to the hadronic
model calculations.

\subsection{Phenomenological parametrization of two-photon exchange effects}
\label{sec:2} Including the TPE effects, $\sigma_R$ of
Eq.~(\ref{diffCr}) takes the following general form
\cite{Guichon03}:
\begin{equation}
\sigma_{R}(Q^2, \varepsilon)= G_M^{2}(Q^2)+
\frac{\varepsilon}{\tau}G_E^{2}(Q^2) + F(Q^2,\varepsilon),
\label{eq:sred2}
\end{equation}
where $F(Q^2,\varepsilon)$ is a real function describing the effect
of $1\gamma\otimes 2\gamma$ interference.  Charge conjugation and
crossing symmetry require  that
  $F$  should, at a fixed
value of $Q^2$, satisfy the following constraint \cite{Rekalo04a},
$F(Q^2,y) = - F(Q^2,-y)$, where
$y=\sqrt{(1-\varepsilon)/(1+\varepsilon)}$. Two  parametrizations
were considered in \cite{Chen07},

\begin{eqnarray}
I: \sigma_{R} &=& G_M^{2}(Q^2)\left(1 + \frac{\varepsilon}{\tau} R^2
\right) + A_1(Q^2) y + A_2(Q^2) y^3,  \label{eq:sred4}\\
II:  \sigma_{R} &=& G_M^{2}(Q^2)\left(1 + \frac{\varepsilon}{\tau}
R^2 \right) + B_1(Q^2) y +B_2(Q^2) y (\ln |y|)^2. \label{eq:sred6}
\end{eqnarray}
Eq. (\ref{eq:sred4}) (parametrization I) is chosen if  $F(Q^2,y)$ is
assumed to be analytic to allow a  Taylor expansion around $y=0$. It
vanishes at $\varepsilon=1\, (y=0)$ and stays finite at
$\varepsilon=0 \,(y=1)$, a feature agrees with the results of the
model calculations of \cite{Blunden03,Chen04}. Eq. (\ref{eq:sred6})
(parametrization II) is allowed if $F(Q^2,y)$ is required to be only
smooth, but not analytic in $0\leq y \leq 1$. The logarithmic or
double-logarithmic functions  are considered since such types of
functions often appear in the loop diagrams \cite{Chen04}. With
functions $A_i(Q^2)$ and $B_i(Q^2)$ assumed to be proportional to
$G_D(Q^2)=1/(1+Q^2/0.71)^2$ and $R$ taken from experiments. It was
found that data for $\sigma_R$ can be fitted with both fits I and II
nicely \cite{Chen07}.

\subsection{$e^+p/e^-p$ ratio and normal asymmetries}
\label{sec:3} The amplitudes for positron-proton (${\it e^+p}$) and
electron-proton scatterings can be written as $T^{(\pm)}=\pm
T_{1\gamma}+T_{2\gamma}$, where  $(\pm)$ correspond to the charge of
positron and electron, and $T_{1\gamma}$ and $T_{2\gamma}$ denote
the scattering amplitudes with $1\gamma$ and $2\gamma$ exchanged,
respectively.  We then have
\begin{eqnarray}
R^{(\pm)}\equiv \frac {\sigma(e^+p)}{\sigma(e^-p)}\simeq 1+
Re\left(\frac{T_{2\gamma}}{T_{1\gamma}}\right),\label{e+/e-}
\end{eqnarray}

\begin{figure}[htbp]
\begin{minipage}[h]{55mm}
\vspace{0pt} \normalsize{ \noindent i.e., measurements of the ratio
of $e^+p$ and $e^-p$ cross sections provide a direct probe of the
real part of the TPE amplitude. Thus measurement of $R^{(\pm)}$
presents a stringent test on any model of TPE. For example, even
though both parametrizations I and II of Eqs. (\ref{eq:sred4}) and
(\ref{eq:sred6}) describe well the experimental data for $\sigma_R$
of comparable quality, their predictions for $R^{(\pm)}$ are very
different as shown in Fig. \ref{fig:3}. However, a recent
measurement of $R^{(\pm)}$ from Novosibirsk  at $Q^2 = 1.6$ GeV$^2$,
$\epsilon= 0.4$, gives $R^{(\pm)} = 1.056 \pm 0.011$
\cite{Nikolenko10} clearly rules out parametrization II. Several
experiments are being performed or planned at JLab  and}
\end{minipage}
\hfill
\begin{minipage}[h]{95mm}
\vspace{0pt} \raggedright
\begin{center}
\includegraphics[width=1.0\textwidth]{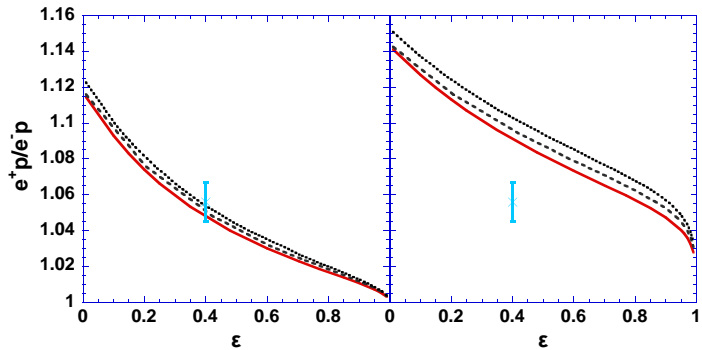}
\end{center}
\caption{Predictions for the ratio $R^{(\pm)}$ of $e^+p$ and $e^-p$
cross sections with parametrizations I (left panel) and II (right
panel) of Eqs. (\ref{eq:sred4}) and (\ref{eq:sred6}), respectively,
as compared with the recent data of
\cite{Nikolenko10}.}\label{fig:3}
\end{minipage}
\end{figure}
\noindent   OLYMPUS to measure $R^{(\pm)}$ for $0.5< Q^2< 2.2 GeV^2$
with $\epsilon > 0.2$. We refer the readers to Ref.
\cite{Arrington11} for more details.

The imaginary part of $T_{2\gamma}$ can be accessed from measurement
of the target normal asymmetry
$A_N\equiv(\sigma^\uparrow-\sigma^\downarrow)/(\sigma^\uparrow-\sigma^\downarrow)$,
where $\sigma^{\uparrow(\downarrow)}$ is the cross section for
unpolarized electrons scattering from a proton target with spin
parallel (antiparallel) to the direction normal to the scattering
plane. To order $\alpha_{em}$, the target normal asymmetry is given
by $A_N=2Im(T_{2\gamma}/T_{1\gamma})$. No data on $A_N$ have been
collected yet. The imaginary part of $T_{2\gamma}$ can also be
accessed by measuring  the beam normal asymmetry for electrons
polarized normal to the scattering plane from unpolarized protons.
Discussions about the experimental and theoretical status of beam
normal asymmetries can be found in \cite{Carlson07}.

\section{Two-boson exchange in parity-violating elastic electron-proton scattering}
Strangeness content in the proton is one of the most intriguing
questions in hadron structure study. Indications on the contribution
of strange quarks to the nucleon properties came from neutrino,
electron deep inelastic scatterings and pion-nucleon sigma term.
Other observables were later suggested, including  double
polarizations in photo- and electroproduction of $\phi$ meson
\cite{Titov97}, and asymmetry $A_{PV}$ in scattering of
longitudinally polarized electrons from unpolarized protons.

 Parity-violating asymmetry
$A_{PV}\equiv(\sigma_R-\sigma_L)/(\sigma_R+\sigma_L)$, where
$\sigma_{R(L)}$ is the cross section with a right-handed
(left-handed)  electron, arises from the interference of weak and
electromagnetic amplitudes. Weak neutral current elastic scattering
is mediated by the $Z$-exchange and measures form factors which are
sensitive to a different linear combination of the three light quark
distributions.  When combined with proton and neutron
electromagnetic form factors and with the use of charge symmetry,
the strange electric and magnetic form factors, $G^s_E$ and $G^s_M$,
can then be determined \cite{Kaplan88}. Specifically, within the
one-boson exchange approximation, $A_2$ in $A_{PV}=A_1+A_2+A_3$  is
given as
\begin{eqnarray}
 A_{2}= a \frac{\epsilon G^{\gamma,p}_{E}G^{s}_{E}
+\tau
G^{\gamma,p}_{M}G^{s}_{M}}{\epsilon(G^{\gamma,p}_{E})^2+\tau(G^{\gamma,p}_{M})^2}=aG^{\gamma,p}_{E}\frac{G^{s}_{E}
+\beta G^{s}_{M}}{\sigma_R(Q^2,\epsilon)}, \label{A2}
\end{eqnarray}
where $\beta=\tau G^{\gamma,p}_{E}/\epsilon G^{\gamma,p}_{E}$,
$a=G_FQ^2/4\pi\alpha\sqrt{2}$ with $\alpha$ the fine structure
constant, while $A_1$ and $A_3$ depend only on nucleon
electromagnetic form factors, Weinberg angle $\theta_W$, and
$G^Z_A$, which can be determined from other measurements. This is a
clean technique to access the charge and magnetization distributions
of the strange quark within nucleons and four experimental programs
SAMPLE, HAPPEX, A4, and G0 \cite{G02010} had been undertaken to
measure this important quantity, which is small and ranges from 0.1
to 100 ppm. It is hence imperative to reduce theoretical uncertainty
on the radiative corrections in order to arrive at a more reliable
interpretation of experiments.

 Leading order radiative corrections to
$A_{PV}$, including the box diagrams in Fig. 4 and other diagrams,
have been extensively studied by Marciano and Sirlin (MS)
\cite{MS83} and widely used in the global

\begin{figure}[htbp]
\begin{minipage}[h]{90mm}
\vspace{0pt} \normalsize{analyses.   Among those corrections, the
interference between $\gamma Z$ exchange ($\gamma ZE$)  of Fig. 4
with  OPE diagram, was evaluated  within the zero momentum transfer
approximation, i.e., $Q^2=0$. As we have
 seen in Sec. 2, it is important to
evaluate  loop diagrams rigorously and accurately. This has been
done in \cite{Zhou07,Nagata09}, and by Tjon and his collaborators
\cite{Tjon08} for both the intereference between $\gamma ZE$ and OPE
diagrams, as well as that between TPE and $Z$-exchange diagrams.
Both groups employed the hadronic model developed in
\cite{Blunden03} and included nucleon and the $\Delta$ in the
intermediate states.

Partonic calculations on the interference  between TPE and
$Z$-exchange, and that between $\gamma ZE$ with OPE have been
carried in \cite{Afanasev05} and \cite{Chen09}, respectively.
 }
\end{minipage}
\hfill
\begin{minipage}[h]{60mm}
\vspace{0pt} \raggedright
\begin{center}
\includegraphics[width=0.85\textwidth]{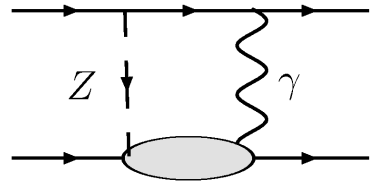}
\end{center}
\caption{ $\gamma Z$ exchange in  electron-proton scattering. The
 cross-box diagram is implied. }
\label{fig:4}       
\end{minipage}
\end{figure}

\
\section{Summary and Outlook}
We have briefly reviewed the advances made in the last decade to
understand the discrepancy in the values of $\mu_p G_E/G_M$ of the
proton as extracted from the traditional Rosenbluth separation of
elastic cross sections and  the new polarization transfer
measurements. Model calculations, both hadronic and partonic,
strongly indicate that the two-photon exchange corrections can at
least account for more the half of discrepancy observed. However, we
are still not yet in the stage of quantifying the two-photon
exchange effects as the theoretical models   still suffered from
uncertainties  such as the on-shell approximation for the vertices
in hadronic calculations and the extrapolations beyond the region of
validity in partonic calculations. Considerable challenge remains on
the theoretical side to obtain a reliable estimate of two-photon
exchange effects.

Just as in the case of nucleon-nucleon interaction, phenomenological
parametrization of the two-photon exchange corrections could be
useful. One of the attempts made in this direction is discussed to
illustrate its potential usefulness. $e^+p/e^-p$ ratio  is known to
probe the real part of the TPE amplitude and can serve as a
stringent test on TPE models. We show that indeed a recent
measurement of $e^+p/e^-p$ ratio  does rule out one of the proposed
phenomenological parametrizations of the two-photon exchange
corrections. More high precision experiments as planned in JLab,
OLYMPUS,  etc. will all be very helpful to further quantify the TPE
effects.

At the end of the day, the final goal of this game is to include TPE
contributions in the analyses of experimental observables and
subsequently extract the form factors which are the aims where we
begin. Such an analysis has been attempted in \cite{Arrington07} and
more will definitely come.

 TPE corrections have been found or speculated
to play an important role in many other observables like hyperfine
splitting in hydrogen, pion form factors, $R_{EM}$ and $R_{SM}$
ratios in $N\rightarrow\Delta$ excitation, as well as in reactions
like electron-nucleus scatterings, deep-inelastic scatterings etc.
We refer the interested readers to the recent excellent reviews
\cite{Carlson07} and \cite{Arrington11} for  details.

Recent studies on the two-boson exchange corrections in the
parity-violating elastic electron-proton scattering, where the
proton strangeness form factors have been extracted, are also
discussed. It is found that the modification incurred in going
beyond the MS approximation  is indeed significant   for some data
  and   the  TBE corrections to $A_{PV}$ depends
strongly on $Q^2$ and $\epsilon$. However, because the effects from
nucleon and $\Delta$ intermediate states tend to cancel out each
other,  the total effects are not large.  It remains to be seen what
role of  other higher resonances might play. This will be a daunting
task in hadronic model calculation since it will introduce many not
well-determined coupling constants and cut-off parameters. Recently,
a dispersion calculation of $\gamma ZE$ corrections to $Q_{weak}$
has been attempted \cite{Gorchtein09}. Many studies along this line
have ensued \cite{Gorchtein11,Blunden11}. It will be very helpful
and welcome if the dispersion relation approach could also be
attempted for the TPE corrections.

%

\begin{acknowledgements}
I would like to dedicate this article to the memory of John A. Tjon.
His deep devotion to  science, pure and simple, has served as a
model to me. I thank  Y.C. Chen, C.W. Kao, K. Nagata and H.Q. Zhou
for collaborations on some of the results presented here. This work
is supported in part by the National Science Council of the Republic
of China (Taiwan) under grant No. NSC99-2112-M002-011.
\end{acknowledgements}



\end{document}